# Spin reorientation behavior in $YMn_{1-x}M_xO_3$ (M = Ti, Fe, Ga; x = 0, 0.1)

**Neetika, A. Das, C L Prajapat[1] and S. S. Meena**

Solid State Physics Division, Bhabha Atomic Research Centre, Mumbai – 400085, India

[1]Technical Physics Division, Bhabha Atomic Research Centre, Mumbai – 400085, India

**Abstract**

The structural and magnetic properties of polycrystalline $YMn_{1-x}M_xO_3$ (M = Ti, Fe, Ga; x = 0, 0.1) have been studied by neutron powder diffraction and magnetic measurements to probe the effect of Mn site doping on the frustration behavior and magnetic structure of these compounds. The compounds are isostructural and crystallize with hexagonal structure in $P6_3cm$ space group. We find that doping with these three ions, $Ti^{4+}$ ($d^0$), $Fe^{3+}$ ($d^5$) and $Ga^{3+}$ ($d^{10}$), influences both the $T_N$ and magnetic structure, unlike other Mn-site dopants reported previously. The magnetic structure of $YMnO_3$ is described by considering a linear combination of irreducible representations $\Gamma 3$ and $\Gamma 4$ below $T_N \sim 75K$ and with decrease in temperature the ratio of $\Gamma 3$ and $\Gamma 4$ changes. The mixing ratio of these two irreducible representations remains constant on lowering of temperature in the Ga doped compounds. The magnetic structure is modified on doping with nonmagnetic ion $Ti^{4+}$ ($d^0$). It is described by the basis vectors of the irreducible representation $\Gamma 2$ with moment $2.3\mu_B$ at 6 K. On doping with $Fe^{3+}$ ($d^5$) the magnetic structure immediately below $T_N$ is explained by considering the $\Gamma 3$ irreducible representation. On further lowering of temperature, a spin reorientation at $\sim 35$ K is observed. Below this temperature, the magnetic structure of $YMn_{0.9}Fe_{0.1}O_3$ is explained by considering $\Gamma 3$ representation with 51% mixing of $\Gamma 4$. The ordered moments are found to be reduced from the expected value for a $Mn^{3+}$ ion in all these compounds indicating the frustrated nature of these compounds. However, the frustration parameter, f is significantly reduced in the case of Ti doped compound with Γ2 representation.

**Introduction**

Multiferroics are materials in which two or all three of the properties viz., ferroelectricity, magnetism and ferroelasticity co-exist simultaneously in a single phase [1] [2]. These are interesting because of their ability to couple two independent phenomena. The $RMnO_3$ compounds with smaller rare-earths ions (R = Ho to Lu and Y) crystallize in the non-centrosymmetric hexagonal space group *P6$_3$cm* [3]. According to Khomskii [4], there are two groups of multiferroics, type-1 and type-2. In type 1 multiferroics, ferroelectricity and magnetism originates from different sources and a weak coupling exists between them which can be seen from the anomalies in the dielectric constant at Néel temperature ($T_N$) [5]. In type-2 multiferroics, the ferroelectric polarization is caused by magnetic ordering and because of this there is strong coupling between them. $YMnO_3$ is a member of type-1 multiferroic compound in which ferroelctricity arises from the cooperative buckling of $MnO_5$ bypiramids leading to displacement of $Y^{3+}$ ions along the c-axis [4]. The ferroelectric transition temperature ($T_{FE}$ ~ 950K) and Néel temperature ($T_N$ ~ 75K) in $YMnO_3$ are well separated from each other. In this compound $Mn^{3+}$ is in 5- fold coordination forming $MnO_5$ trigonal bypiramids which forms a quasi - 2D triangular network in ab plane. This leads to geometrical frustration and it is evidenced by large ratio of Curie-Weiss to Neel temperatures ($\theta/T_N$). Several neutron diffraction measurements have been carried out to determine the magnetic structure of $YMnO_3$.Two magnetic structures of α-type (the $\Gamma_3$ representation of space group *P6$_3$cm* ) and β type ( the $\Gamma_1$ representation of space group *P6$_3$cm* ) has been suggested for $YMnO_3$ by Bertaut et. al [6]. And according to Munoz et. al [7], the β - type structure describes better the magnetic structure of $YMnO_3$. However, from neutron diffraction experiment it is not possible to distinguish between these two representations $\Gamma_1$ and $\Gamma_3$. The magnetic structures which were indistinguishable by neutron diffraction could be distinguished by non-linear optical spectroscopy [8]. In this study the ground state of YMnO$_3$ has been explained by Γ3 (*P6$_3$cm*) representation. In *P6$_3$cm* structure there are two types of interactions, between nearest neighbor (NN) and next nearest neighbor (next-NN). But the behavior of NN interactions and next-NN interactions obey different rules. The strength of exchange coupling between nearest neighbour is directly related to bond lengths but in the case of next-NN interactions a simple analysis of bond lengths is not applicable since these interactions involve intermediate paths. For the next-NN interactions between planes, there are two equivalent strong bonds and one weak bond. So to minimize the energy of the system, there should be a FM coupling in the weak bond and this FM coupling is present only in Γ3 and not in Γ1. By taking into account these next- NN interactions it has been shown, theoretically, that Γ3 irreducible representation is stabilized in YMnO$_3$ [9].

Divalent substitutions (Ca, Sr) at the Y site over a wide composition regime show a variety of structural and magnetic phases [10, 11, 12, 13]. Doping trivalent Er at Y site in YMnO$_3$ the magnetic structure changes from Γ1 representation in YMnO$_3$ to Γ2 representation in ErMnO$_3$ [14] and the system becomes less frustrated with no significant reduction in ordered moment is observed. While, in the case of doping with Lu the frustration parameter remains constant and magnetic structure for all the doped samples has been explained by a mixture of Γ3 and Γ4 but the angle [15] between the moments and crystallographic axes is found to change form 10º for YMnO$_3$ to 83.6º for LuMnO$_3$. In Y$_{0.8}$Tb$_{0.2}$MnO$_3$, in addition to AFM order of the Mn$^{3+}$ ions at T ~ 71 K additional transition at T ~ 23 K is observed which corresponds to the Mn spin reorientation [16].

The frustrated behavior in these compounds has been also found to be influenced by doping at the Mn site. Spin glass state has been reported in Mn rich hexagonal manganite YMn$_{1+x}$O$_3$ (0

≤ x ≤ 0.15) and in $YMn_{1-x}Cr_xO_3$ (0 ≤ x ≤ 0.1) [17] [18]. With Fe doping at Mn site in $YMnO_3$ a single phase has been obtained for x ≤ 0.3 [19]. The transition temperature decreases to 60K with 20% Fe doping and reduction in effective moments has also been observed [20]. In $YMn_{1-x}Ti_xO_3$, a structural phase transition from hexagonal ($P6_3cm$) to rhombohedral ($R3c$) is observed around x = 0.2 and the Curie-Weiss temperature is found to decrease with increase in Ti concentration indicating the suppression of average antiferromagnetic interactions [21]. By wet chemistry techniques single-phase hexagonal type solid solution has been formed for Cu doping at Mn site in $YMnO_3$ and self doping at the Y-site [22]. $Cu^{2+}$ doping results in partial transformation of the $Mn^{3+}$ into $Mn^{4+}$, and this introduces weak ferromagnetic interactions $Mn^{3+}$- $Mn^{4+}$. Doping with 10% Al, Ru, and Zn a decrease in Mn moment and a slight decrease in $T_N$ has been observed. In this study the magnetic structure of $YMnO_3$ has been explained by taking a mixture of $\Gamma_3$ and 18% of $\Gamma_4$ and with doping of 10% Al, Ru, Zn the mixing ratio of $\Gamma_3$ & $\Gamma_4$ is found to be modified [23]. A spin reorientation of Mn magnetic moments has been observed in YMnO3 under high pressure (5GPa) and a change in magnetic ground state has been seen which can be described by a combination of $\Gamma1$ and $\Gamma2$ irreducible representation [24]. Magnetoelastic coupling also has been observed in this compound [25]. From Raman studies spin phonon coupling has been observed in this compound below $T_N$ [26, 27]. The doping experiments at the Y-site and Mn site and external pressure experiments suggest that the magnetic structure of $YMnO_3$ is subject to alterations by these. The bond lengths and bond angles play a crucial role in stabilization of the magnetic structures as some of these studies show [24]. In the present work we report the effects of $Ti^{4+}$ ($d^0$), $Fe^{3+}$ ($d^5$) and $Ga^{3+}$ ($d^{10}$), doping on the structural and magnetic structure of $YMnO_3$. The dopants have been chosen to study the effect of interaction of the filled and unfilled orbital on the magnetic structure. We find that these three dopants affect the magnetic structure of this compound in contrasting manner. The compositions were limited to 10% doping to remain in the isostructural phase.

**Experimental Details**

Polycrystalline samples of $YMn_{1-x}M_xO_3$ (M = Ti, Fe, Ga; x = 0, 0.1) were synthesized through a solid state reaction by heating stoichiometric quantities of $Y_2O_3$, $MnO_2$, $TiO_2$, $Ga_2O_3$ and $Fe_2O_3$ in air at 1200°C for 90 hrs with several intermediate grindings. Phase identification of these samples was done by x-ray powder diffraction recorded on a Rigaku diffractometer, using Cu Kα radiation in the angular range $10^o \leq 2\theta \leq 70^o$ at room temperature. The magnetization measurements, both in zero field cooled (ZFC) and field-cooled warming (FCW) conditions, were carried out by using superconducting quantum interference design (SQUID) magnetometer. The neutron diffraction patterns were recorded on a multi-PSD-based powder diffractometer ( λ = 1.2443Å ) at the Dhruva reactor, Bhabha Atomic Research Centre, Mumbai between 6K and 300K in the angular range $5^o \leq 2\theta \leq 140^o$. The neutron diffraction patterns were refined using the FULLPROF program [28]. A Mössbauer spectrum of sample was recorded using a Mössbauer spectrometer (Nucleonix Systems Pvt. Ltd., Hyderabad, India) operated in constant acceleration mode (triangular wave) in transmission geometry at room temperature. The source employed was Co-57 in Rh matrix of strength 50 mCi. The calibration of the velocity scale was done by using an enriched a-$^{57}$Fe metal foil. The line width (inner) of calibration spectra was 0.23 mm/s. The Mössbauer spectrum was fitted with a least square fit (MOSFIT) program assuming Lorentzian line shape.

**Results and Discussion**

The x-ray diffraction patterns and the Rietveld refinement of the studied polycrystalline samples YMn$_{1-x}$M$_x$O$_3$ (M = Ti, Fe, Ga; x = 0, 0.1) are shown in figure 1. It indicates that all the samples are isostructural and crystallizes in hexagonal phase (space group *P6$_3$cm*). The lattice parameters and unit cell volume obtained from x-ray diffraction of YMnO$_3$ agrees with the earlier reported values [7]. With Fe doping both the lattice parameters a and c increases and the values obtained are similar to the values reported by Zaghrioui et al.[20] while for Ga and Ti doping a increases and c decreases. In the case of Ga doping our result of lower c/a is different from the behavior reported in single crystal studies of YMn$_{1-x}$Ga$_x$O$_3$ [29, 30], where Ga doping is found to increase the ratio of c/a. The variation of the cell parameters in Ti doped sample are similar to those reported previously where the decrease of c parameter has been ascribed to decrease of the tilting of MnO$_5$ bypiramids [21]. The increase in the cell parameters of Fe doped sample is understood as follows. In trigonal bypiramidal geometry, the d- levels are split into two doublets (d$_{xz}$, d$_{yz}$, d$_{x^2-y^2}$ and d$_{xy}$) and one singlet (d$_z^2$). The four d- electrons of Mn$^{3+}$ occupy lowest lying doublets and no electron is present in d$_z^2$ orbital. Whereas, Fe$^{3+}$ doping introduces one more electron in the d$_z^2$ orbital resulting in elongation of the c –axis as argued previously in Fe-doped samples [19, 31]. However, it is difficult to explain the behavior of the lattice parameter a in view of the similar ionic radii of Mn$^{3+}$ and Fe$^{3+}$ in five fold coordination (0.58 Å). The results of refinement of neutron diffraction data taken at room and low temperature for all the studied compounds are included in table 1. For all the studied samples, the lattice parameter a increases and c decreases with increase in temperature. This behavior is reported to persist until about 1270 K above which the cell parameters tend to be roughly constant [32]. The negative thermal expansion of c parameter is explained by the reduction of tilting of MnO$_5$ bypiramids along with the buckling of Y-planes [33]. Variation of cell parameters for YMnO$_3$ with temperature is shown in fig 2(a). The behavior is the same in all the three doped compounds. The temperature dependence of volume has been fitted to Debye - Grüneisen equation [34]. In the Grüneisen approximation, the temperature dependence of volume is described by, V(T) = γ U(T)/B$_0$ + V$_0$, where γ, B$_0$, and V$_0$ are the Grüneisen parameter, bulk modulus and volume, respectively, at T = 0 K. In the Debye approximation, internal energy U (T) is given by,

$$U(T) = 9Nk_BT\left(\frac{T}{\theta_D}\right)^3 \int_0^{\frac{\theta_D}{T}} \frac{x^3}{e^x - 1}dx,$$

where $x = \frac{h\nu}{k_BT}$, N is the number of atoms in the unit cell, k$_B$ is the Boltzmann's constant [35].

Fig 2(b) shows the temperature variation of unit cell volume of YMnO$_3$ and the corresponding fit to the Debye - Grüneisen equation (Grüneisen approximation is considered by using Debye model ). The unit cell volume contracts and below T$_N$ the fit shows a clear deviation from the behavior expected from the temperature dependence of volume of a non magnetic compound described by the above equation. This anomalous contraction of unit cell volume below T$_N$ is an evidence of magnetoelastic coupling in YMnO$_3$. Similar magnetoelastic coupling has been previously shown in this compound [25]. The refined Mn-O and Y$_{1,2}$-O bond lengths are also shown in Table 1. The apical Mn-O1 and Mn-O2 bond lengths are smaller than the planar Mn-O3 and Mn-O4 bond lengths for all the studied samples and this is consistent with the previous studies [33,36]. The tilting and buckling of MnO$_5$ bipyramid are important lattice distortion parameters, which are expected to change with doping at Y-site or Mn- site. The tilting angle (α) is defined by the angle between the O1-O2 axis of the MnO$_5$ bipyramid and c axis and the buckling is represented by the angle β between the O3-O4-O4 plane and c axis [36]. The experimentally obtained values of the tilting and buckling angles for YMnO$_3$ are in close agreement with the theoretically obtained values [37]. Substitution of Mn with Ti$^{4+}$ (d$^0$), Fe$^{3+}$ (d$^5$) and Ga$^{3+}$ (d$^{10}$) changes the tilting and buckling of MnO$_5$ trigonal bipyramids as shown in figure 3(a) and 3(b). Substituting Mn with Ti and Fe reduces the tilting and buckling whereas Ga increases the tilting of MnO$_5$ trigonal bipyramids which is in good accordance with the theoretically calculated values [37]. We find that decrease in buckling, as in the case of Ti doped sample, leads to lowering of frustration.

The temperature dependence of magnetization M(T) under an applied magnetic field of 0.1 T for all the samples is shown in figure 4(a). The magnetization increases on lowering of temperature and a distinct anomaly at the transition temperature is observed only in the case of Fe. Similar absence of anomaly at the T$_N$ has been reported previously in YMnO$_3$ [7]. In the case of Ti doped sample the enhancement of M at low temperatures is higher as compared to other samples. We attribute this behavior to the small out of plane ferromagnetic component in the Γ2 magnetic structure observed in this sample (discussed later). The inverse magnetic susceptibility versus temperature curve is shown in figure 4(b). It exhibits a large curvature extending to high temperatures and therefore, we found that it could not be fitted to the Curie – Weiss law. Evidence of short range ordering in the proximity of the T$_N$ has been reported in the parent compound [38] and explains the departure from CW behavior in the vicinity of T$_N$. The doping is found to extend the temperature range above T$_N$ where the short range ordering persists. Diffuse scattering studies in half doped manganites, show evidence of magnetic short range ordering far above the magnetic ordering temperature [39]. We found a better description of the paramagnetic susceptibility data by fitting the magnetic susceptibility to the modified Curie – Weiss law, given by $\chi = \chi_0 + C/T-\theta_{CW}$, where $\chi_0$, C and $\theta_{CW}$ are the temperature independent susceptibility, Curie constant and Curie-Weiss temperature, respectively. The values of $\chi_0$, the effective Mn moments ($\mu_{eff}$) and the paramagnetic temperature ($\theta_{CW}$) could be obtained from this fit and is summarized in Table 1. The magnetic susceptibility follows a modified Curie – Weiss behavior above 190 K for YMnO$_3$, above 175 K for YMn$_{0.9}$Ti$_{0.1}$O$_3$, above 185K for YMn$_{0.9}$Fe$_{0.1}$O$_3$ and above 225 K for YMn$_{0.9}$Ga$_{0.1}$O$_3$ as shown in fig 4(b). The values of $\theta_{CW}$ and $\mu_{eff}$ obtained for YMnO$_3$ are -421K and 4.98 $\mu_B$, respectively. With doping of Ti a pronounced reduction in $\theta_{CW}$ (-119K) is observed while in the case of Fe and Ga the reduction is marginal (Table I). The effective paramagnetic moment is 4.30 $\mu_B$, 4.45 $\mu_B$ and 4.10 $\mu_B$ for YMn$_{0.9}$Ti$_{0.1}$O$_3$, YMn$_{0.9}$Fe$_{0.1}$O$_3$ and YMn$_{0.9}$Ga$_{0.1}$O$_3$, respectively. Theoretically, $\mu_{eff}$ is calculated as, $\mu_{eff}^{cal} = \sqrt{x\mu_{eff}^2(M)+(1-x)\mu_{eff}^2(Mn^{3+})}$ , where $\mu_{eff}$ for Mn$^{3+}$ ( S = 2 ) is 4.89 $\mu_B$. According to this equation, $\mu_{eff}^{cal}$ values for Ti, Fe and Ga doped compounds are 4.65 $\mu_B$, 5.0 $\mu_B$ and 4.65 $\mu_B$ respectively. With 10% doping of Ti, Fe and Ga at Mn site of YMnO$_3$, Curie - Weiss temperature as well as effective moment reduces. Substitution of Ti$^{4+}$ at Mn$^{3+}$ site can lead to the formation of Mn$^{2+}$ by introduction of electrons in the system, assuming stoichiometric oxygen. But in magnetization measurement we observe reduction in effective moment value which is inconsistent with the presence of Mn$^{2+}$ in this sample. This suggests that there is a change in the Oxygen stoichiometry in the sample. Previous studies of magnetic susceptibility on samples prepared under reducing atmosphere have shown that at a given temperature the paramagnetic susceptibility decreases for the reduced sample which again suggests the absence of Mn$^{2+}$ [21]. So doping of Ti$^{4+}$ at Mn site changes the stoichiometry of oxygen in these samples though, we believe, the change is too small to influence the magnetic structure of the compound. The magnetic structure of YMnO$_{3-\delta}$ ($\delta \sim 0.29$) has been found to be the same as that of YMnO$_3$ albeit, with a different tilt angle [40]. Hence, the oxygen non-stoichiometry does not appear to influence the spin structure, though an enhancement in the transition temperature has been observed in oxygen non- stoichiometric compound $YMnO_{3-\delta}$ [41]. The decrease of effective moment ($\mu_{eff}$) in Fe doped sample is unexpected since $Fe^{3+}$ has one more unpaired electron than $Mn^{3+}$. The decrease of $\mu_{eff}$ in this sample could be explained by the presence of $Fe^{2+}$. But Mössbauer spectroscopy excludes the presence of $Fe^{2+}$ (discussed below). Therefore, this behavior is explained by considering a competition between ferromagnetic Fe-O-Mn interactions and antiferromagnetic interactions Mn-O-Mn and Fe-O-Fe [20]. The values of $\theta_{CW}$ and $\mu_{eff}$ for all these samples are given in table 1. Using these values of $\theta_{CW}$ we have estimated the exchange integral J = 3.0 meV between the nearest Mn neighbors using the expression $\theta_{CW} = -z\,J\,S(S+1)/3$ [38, 42], where S = 2 for $Mn^{3+}$, z = 6 is the number of nearest neighbors and $\theta_{CW}$ is Curie - Weiss temperature. Doping at Mn site with $Ti^{4+}$ ($d^0$), $Fe^{3+}$ ($d^5$) and $Ga^{3+}$ ($d^{10}$) reduces this exchange integral to 0.93 meV, 2.29 meV and 2.9 meV, respectively. Since Curie – Weiss temperature ($\theta_{CW}$) is a measure of AFM coupling strength between Mn ions, the results suggest that doping with Ti suppress the AFM interaction. The extrapolated paramagnetic temperatures ($\theta_{CW}$) in $YMn_{0.9}M_{0.1}O_3$ (M = Ga, Ti and Fe) are much higher than $T_N$. This difference in the values of $\theta_{CW}$ and $T_N$ is an evidence of the magnetic frustration in these compounds. This is expressed by frustration parameter, f = $|\theta_{CW}|$ / $T_N$ = 5.6 for $YMnO_3$ while it becomes 2.2, 5.6 and 6.9 for $YMn_{0.9}Ti_{0.1}O_3$, $YMn_{0.9}Ga_{0.1}O_3$ and $YMn_{0.9}Fe_{0.1}O_3$, respectively. Among the three dopants a significant reduction in the frustration parameter is observed in the case of Ti, while it remains same for Fe and increases in the case Ga doped sample. Similar behavior has been observed previously in Ga doped $YMnO_3$ [30]. Reduction of frustration has been seen in other doped samples, for e.g. on doping with non magnetic ions such as Ru, Al, and Zn at Mn site, frustration parameter reduces. Doping Er at Y-site is also found to reduce frustration in these systems [14].

**Mössbauer spectrometry**

The Mössbauer spectrum of $YMn_{0.9}Fe_{0.1}O_3$ sample was recorded at room temperature to confirm the oxidation state of Fe and is shown in figure 5. The spectrum is fitted with three symmetric doublets, indicative of three different chemical environments around Fe atom and all the Fe being paramagnetic. In this structure Mn occupies only one crystallographic site in the unit cell. The three different sites for Fe observed in Mössbauer studies arises due to the random distribution of Fe in lattice having different numbers of $Mn^{3+}$ ion near neighbors. These three different chemical environments around Fe atom could be understood as follows. In this structure, Mn (or Fe) site has six in plane nearest neighbors. The system consists of 10% Fe and the rest is Mn. So, the probability of having a Fe atom in the nearest neighbor site is 0.1 (10%) and the probability of Mn is 0.9 (90%). Thus for a given Fe ion, there is 53% chance of having all Mn atoms as nearest neighbors ( no Fe atoms), 35% chance of having only one Fe atom (five Mn atoms) and 10% chance of having two Fe atoms. For more than two Fe atoms probability reduces significantly. The relative area for each doublet is in agreement with the calculated probabilities. However, our results differ from the earlier reported Mössbauer study on the same composition where the Mössbauer data was fitted with two doublets and was ascribed to two different chemical environments around the Fe atom [20]. The hyperfine parameters i.e. isomer shift ($\delta$), quadrupole splitting ($\Delta E_Q$) and line widths ($\Gamma$) obtained from the fit are included in Table 2. The value of isomer shift for all the three doublets is ~ 0.30 mm/s at room temperature and is in agreement with previously reported values for this compound [20]. These value corresponds to the presence of $Fe^{3+}$ in high spin state (S = 5/2) [43]. Quadrupole splitting ($\Delta E_Q$) arises due to interaction between electric quadrupole moment of the nucleus and surrounding electric field gradient and thus give relevant information about the charge symmetry around the nucleus. The large value of quadrupole splitting in this sample is thus attributed to the large distortion of Fe site.

**Magnetic Structure**

Neutron diffraction pattern has been recorded for all the samples at several temperatures below 300K. The diffraction data at room temperature show that the samples are isostructural and crystallizes in the hexagonal structure confirming the x-ray diffraction results and no structural transition is observed on lowering of temperature. Thermal evolution of neutron data for YMnO$_3$ shows that magnetic reflections (100) (101) and (102) together with fundamental reflections gain in intensity with decrease in temperature below 75K. These reflections have been indexed using identical dimensions for the magnetic and chemical cell (propagation vector, k = 0) in P-1 space group. Munoz et al. [7] in their study of YMnO$_3$ has shown that there are six representations ($\Gamma$1 – $\Gamma$6) that are possible in this hexagonal structure. We have used the Sarah program [44] to carry out the representation analysis and obtain the basis vectors for each of these representations.

In YMnO$_3$, the magnetic structure can be described either by $\Gamma$1 or $\Gamma$3 irreducible representation (IR), since for both these IRs nearly identical diffraction intensities are observed. In $\Gamma$1 representation, the magnetic moment has a component in the xy plane and the spins are oriented perpendicular to the x-axis. The z = 0 and z = ½ layers are antiferromagnetically coupled in $\Gamma$1. In $\Gamma$3 representation, the magnetic moment has a component in xy plane and a component along the z axis. The coupling is ferromagnetic for both, the component in xy plane and for the component along the z axis. In the theoretical studies it has been shown that $\Gamma$3 irreducible representation is stabilized by the next – NN interactions and Γ1 irreducible representation is not an appropriate description for the magnetic structure of YMnO$_3$ [9]. Therefore, we had analyzed our neutron diffraction data by considering Γ3 IR alone. But the intensity of (101) reflection does not match well by taking Γ3 IR alone. Then we reanalyzed our neutron diffraction data by considering mixed representations Γ3 + Γ2 and Γ3 + Γ4 irreducible representations. Though the fits are equally good in these two cases, the magnetic structure obtained by considering linear combination of Γ3+Γ2 irreducible representations is not realistic, since ordered moment for Mn are different in z = 0 and z = ½ plane. So, we conclude that the ground state of YMnO$_3$ is best described by a linear combination of Γ3 and Γ4 and this result is in agreement with those reported earlier [23]. Fig 6 shows Reitveld refinement of neutron diffraction data for YMnO$_3$ at 300 K and 6 K. The value of the moment on Mn is 3.24 $\mu_B$ is lower than the expected value 4 $\mu_B$ for Mn$^{3+}$ (S = 2). The observed lower value of Mn moment as compared to the expected value is ascribed to frustration. Similar magnetic ground state and reduction in magnetic moment has also been observed on this compound in previous studies [7, 23]. In this structure the moments are in the a-b plane and tilted away from the crystallographic a-axis by a tilting angle (φ). The angle (φ) in case of YMnO$_3$ at 6K is 11.8° i.e. moment on Mn is inclined at 11.8° to the a axis. This value is close to that reported in the magnetic structure studies by Brown and Chatterji in which Mn moments are aligned at 11.1° to the [100] plane [45] and by Park et al. [23] in which the angle (φ) is 10.3°. The tilting angle increases with increase in temperature (inset of figure 6). This behavior of the tilt angle as a function of temperature is different than that observed in other compounds in the series RMnO$_3$, possibly arising from the absence of moment at the rare earth site. In the case of HoMnO$_3$, below T = 78.5 K the magnetic structure is explained by Γ2 irreducible representation with moments parallel to [100] axes. With decrease in temperature below T = 44.6 K moments rotate in basal plane in such a way that below T = 38.8 K the magnetic structure is explained by Γ1 with moments perpendicular to [100] axes [46]. In ScMnO$_3$, the angle φ changes to 54º from 0º as the temperature reduces from 89.4 K to 1.8 K [7]. Magnetic order in ErMnO$_3$ and in YbMnO$_3$ is explained by Γ2 or Γ4 without any spin reorientation behavior as seen in HoMnO$_3$. The thermal evolution of magnetic moments for YMnO$_3$ is shown in inset of fig 6. We have applied molecular field analysis to describe the thermal variation of Mn magnetic moment. Using the Brillouin description for reduced magnetization, $m_{Mn} = m_{sat}(T)B_2(x)$, where $x = \dfrac{mN\lambda(gS\mu_B)^2}{k_B T}$, $m_{sat}$ = 3.24 μ$_B$, S = 2 for Mn$^{3+}$, T$_N$ = 75 K, and molecular field constant $\lambda = \dfrac{3k_B T_c}{g^2 S(S+1)\mu_B^2}$ = 13.9 T/μ$_B$ [47], we obtained a good fit to the experimental data, as shown in inset of fig 5. Here we have used T$_N$ as a parameter in the absence of neutron diffraction data at smaller intervals of temperature. The mean field approximation is found to describe well the magnetism in this frustrated compound.

Doping with Ti$^{4+}$ (d$^0$), Fe$^{3+}$ (d$^5$) and Ga$^{3+}$ (d$^{10}$) at Mn site in YMnO$_3$ reduces T$_N$ significantly. Fig 7(a) shows the evolution of integrated intensity of (100) magnetic reflection for parent sample, Ga and Fe doped samples as a function of temperature. The evolution of integrated intensity of (101) magnetic reflection for YMn$_{1-x}$M$_x$O$_3$ (M = Ti, Fe, Ga; x = 0, 0.1) is shown in fig 7(b). In comparison, doping with Al, Ru, Zn at Mn site is found to show a very small reduction in T$_N$ [23]. This suggests that the present dopants strongly influence the Mn-Mn interactions.

On doping with Ti$^{4+}$ in YMnO$_3$ at Mn site, T$_N$ reduces to 55 K. The (100) reflection which is predominately magnetic is absent in this sample but the (101) magnetic reflection is evident as shown in figure 8. The magnetic structure in this case is could be fitted by both the irreducible representation Γ2 and Γ4. In Γ2 representation, the magnetic moment has a component in xy plane and a component along the z axis. The coupling is antiferromagnetic for the component in xy plane and ferromagnetic for the component along the z axis. In Γ4 representation the spins are in the xy plane and are oriented perpendicular to the a-axis. The z=0 and z= ½ layers are ferromagnetically coupled in Γ4. In this sample, M(T) shows a significant increase in M on lowering of temperature as compared to other samples in the series. We measured the field dependence of magnetization (M-H curves) at 5 K for $YMnO_3$ and for $YMn_{0.9}Ti_{0.1}O_3$. For $YMnO_3$, the magnetization curve do not show any hysteresis phenomenon or saturation magnetization but for $YMn_{0.9}Ti_{0.1}O_3$ M(H) exhibits a hysteresis (shown in the inset to Figure 7). These two observations suggest the presence of a small ferromagnetic component in the magnetization of Ti doped sample. A ferromagnetic component is supported in Γ2 and not in Γ4. Therefore, we have chosen Γ2. The Rietveld refinement of neutron diffraction data for $YMn_{0.9}Ti_{0.1}O_3$ at 6 K is shown in fig 8. Thus, Ti doping leads to the modified magnetic structure for $YMn_{0.9}Ti_{0.1}O_3$ and is given by irreducible representation Γ2 with moment on Mn is 2.3$\mu_B$ at 6 K and is oriented along the a axis. The ferromagnetic component of the moment is lower than that we can detect using neutron diffraction. Therefore, we do not find the out of plane component of moment in the analysis of neutron diffraction data though signatures of this we can find in the magnetization data. The occurrence of weak ferromagnetic component in Ti doped samples can be explained by considering Dzyaloshinskii-Moriya (D-M) type interactions [48]. The thermal evolution of the moments is described by the molecular field analysis using λ = 11.2 T/$\mu_B$, $T_N$ = 55 K and S = 1.9. We obtained a good fit to the experimental data as shown in inset of figure 8.

As against the previous two samples in the Fe doped sample spin reorientation as a function of temperature is observed. The neutron diffraction patterns for $YMn_{0.9}Fe_{0.1}O_3$ at T = 55 K is shown in fig 9. The magnetic phase observed immediately below $T_N \sim 55$ K, is described by considering irreducible representation $\Gamma 3$. This compound undergoes a second magnetic transition at T ~ 35 K, where the magnetic reflection (101) begins to be observed. In comparison to $YMnO_3$, where the intensity of the (100) and (101) peaks are quite different, in $YMn_{0.9}Fe_{0.1}O_3$ at 6 K the intensity of these two peaks is almost same as shown in fig 10. So Fe doping at Mn site leads to spin reorientation of Mn magnetic moments and this modified magnetic structure can be described as a linear combination of irreducible representation $\Gamma 3 + \Gamma 4$ with different mixing ratio of these two representations. The magnetic structure of $YMnO_3$ at 6 K is described by $\Gamma 3$ with 26% mixing of the $\Gamma 4$ representation. With Fe doping this mixing ratio changes to 51% and the angle ($\varphi$) changes from $11.8°$ for $YMnO_3$ to $28°$ for $YMn_{0.9}Fe_{0.1}O_3$ at 6K. The change in magnetic ground state of $LuMnO_3$ ($\Gamma 4$) from $YMnO_3$ ($\Gamma 3$) has been ascribed to the behavior of single ion anisotropy which is correlated with the distortions of triangular lattice [9]. The presence of $d_z^2$ orbital in Fe doped $YMnO_3$ influences the anisotropy of the system leading to reorientation of the spins. In earlier studies of Fe doping in $YMnO_3$ it has been seen that Fe doping introduces more magnetic anisotropy in the system [20]. The moment on Mn reduces to $2.93\mu_B$. In case of $YMnO_3$, $\varphi$ angle increases with increase in T whereas for Fe-doped samples $\varphi$ angle decreases with increase in temperature. Spin reorientation behavior has also been observed in $HoMnO_3$ below T = 44.6 K [42] and in $ScMnO_3$ where the spin reorientation changes the magnetic symmetry from $\Gamma 2$ to $\Gamma 1 + \Gamma 2$ at low temperature [7]. The thermal variation in Mn magnetic moments for $YMn_{0.9}Fe_{0.1}O_3$ derived from neutron data is shown in inset of fig 10. We calculate these thermal variations by applying molecular field model. For these doped sample, we obtained $\lambda = 10.7$ T/$\mu_B$ by taking $T_N = 60$ K and S = 2.05.

We find that the representation remains same on doping with Ga in YMnO$_3$ albeit with a decrease in the value of moment to 2.07 $\mu_B$ at 6 K. The tilting angle for this compound is 16.4º and this remain almost same for all temperatures below $T_N$. The mixing ratio of these two irreducible representation can be related to the ratio of intensities of (100) and (101) peaks. Since (100) is a pure magnetic peak and it is completely absent for $\Gamma 4$ irreducible representation. For Ga doped compound the ratio of intensities of (100) and (101) peaks remain almost same as that of the parent compound. So, doping at Mn site with Ga does not change the mixing ratio of $\Gamma 3$ and $\Gamma 4$ though it reduces the $T_N$. In comparison, doping with nonmagnetic ions Ru, Al and Zn at Mn site changes the mixing ratio of these two irreducible representations.

**Conclusion**

In the present work, we have studied the effect of Mn-site doping on the magnetic structure of YMnO$_3$ and find their varied influences. Polycrystalline samples of YMn$_{1-x}$M$_x$O$_3$ (M = Ga, Ti, Fe; x = 0, 0.1) crystallize in hexagonal structure (*P6$_3$cm* space group). A significant reduction is observed in $T_N$ on doping with 10% Ti$^{4+}$ (d$^0$), Fe$^{3+}$ (d$^5$) and Ga$^{3+}$ (d$^{10}$). The magnetic structure of YMnO$_3$ is described by considering a linear combination of irreducible representations $\Gamma 3$ and $\Gamma 4$ below $T_N \sim 75$K and with decrease in temperature the ratio of $\Gamma 3$ and $\Gamma 4$ changes. The $T_N$ of the Ga doped compound reduces whereas the magnetic structure is described by the same set of IR albeit with a reduced value of ordered moment. On doping with Ti the magnetic structure is described by IR $\Gamma 2$. In this structure an out of plane weak ferromagnetic component appears which influences the M(T,H) data. Doping with Ti changes the magnetic ground state of YMnO$_3$ and system evolves to less frustrated systems. On doping with Fe$^{3+}$ (d$^5$) the magnetic phase which appears immediately below $T_N$ is explained by considering the $\Gamma 3$ irreducible representation and undergoes a second magnetic transition around ~ 35 K, corresponding to a spin reorientation. Below this temperature the modified magnetic structure of $YMn_{0.9}Fe_{0.1}O_3$ is explained by a linear combination of $\Gamma 3$ and $\Gamma 4$ with reduced moment $2.93\mu_B$ at 6 K. The absence of d orbital as in the case of Ti leads to loss of frustration behavior whereas the fully filled d orbital in Ga enhances the frustration in the system.

**Acknowledgement**

We thank the referee for correcting our analysis of the Mössbauer data.

**Figure Captions:**

**Fig 1:** Room temperature powder X- ray diffraction patterns of $YMn_{1-x}M_xO_3$ (M = Ti, Fe, Ga; x = 0, 0.1). Open circles are observed data points. The solid line represents the Rietveld refinement.

**Fig 2 (a):** Temperature variation of lattice parameters a and c. (b) Temperature variation of unit cell volume and solid line represent a fit to Debye - Grüneisen equation at high temperature above $T_N$, and then this extrapolated to the lowest temperature below $T_N$.

**Fig 3 (a):** Tilting angle for all doped sample at 6K. The dashed line denotes the tilting of $YMnO_3$ (4.5º). (b) Buckling angle of $MnO_5$ trigonal bypiramids for all doped samples at 6K. The dashed line denotes the buckling in $YMnO_3$.

**Fig 4 (a):** The zero field-cooled (ZFC) magnetization (M) versus temperature (T) in field of H = 0.1T for $YMn_{1-x}M_xO_3$ (M = Ti, Fe, Ga; x = 0, 0.1). (b) Shows the inverse of susceptibility as a function of temperature and modified Curie- Weiss fit (solid line).

**Fig 5:** Room temperature Mössbauer spectrum of $YMn_{0.9}Fe_{0.1}O_3$.

**Fig 6**: The observed (symbols) and calculated (line) neutron diffraction pattern for $YMnO_3$ compound at T = 6 K and 300 K. Lower solid line is the difference between observed and calculated pattern. The first row of tick marks indicates the position of nuclear Bragg peaks and second row indicate the position of magnetic Bragg peaks. Inset (a) shows the variation of tilting angle (φ) as a function of temperature. Inset (b) shows the variation of magnetic moment as a function of temperature.

**Fig 7 (a):** Integrated intensity of (100) magnetic reflection for $YMn_{1-x}M_xO_3$ (M = Ga, Fe; x= 0, 0.1) as a function of temperature. (b) Integrated intensity of (101) magnetic reflection for $YMn_{1-x}M_xO_3$ (M = Ti, Fe, Ga; x = 0, 0.1) as a function of temperature.

**Fig 8:** The observed (symbols) and calculated (line) neutron diffraction pattern for $YMn_{0.9}Ti_{0.1}O_3$ compound at T = 6 K. Lower solid line is the difference between observed and calculated pattern. The first row of tick marks indicates the position of nuclear Bragg peaks and second row indicate the position of magnetic Bragg peaks. Inset (a) shows the M-H curve at T = 5 K. Inset (b) shows the variation of magnetic moment as a function of temperature.

**Fig 9:** The observed (symbols) and calculated (line) neutron diffraction pattern for $YMn_{0.9}Fe_{0.1}O_3$ compound at T = 55 K. Lower solid line is the difference between observed and calculated pattern. The first row of tick marks indicates the position of nuclear Bragg peaks and second row indicate the position of magnetic Bragg peaks.

**Fig 10:** The observed (symbols) and calculated (line) neutron diffraction pattern for $YMn_{0.9}Fe_{0.1}O_3$ compound at T = 6 K. Lower solid line is the difference between observed and calculated pattern. The first row of tick marks indicates the position of nuclear Bragg peaks and second row indicate the position of magnetic Bragg peaks. Inset (a) shows the variation of tilting angle ($\varphi$) as a function of temperature. Inset (b) shows the variation of magnetic moment as a function of temperature.

**Table Captions:**

**Table 1:** Results of Rietveld refinement of neutron diffraction pattern at 6K and 300K, Curie – Weiss fit parameters, geometrical frustration parameter, and transition temperature for $YMn_{1-x}M_xO_3$ (M = Ti, Fe, Ga; x = 0, 0.1).

**Table 2:** Mössbauer parameters at room temperature for $YMn_{0.9}Fe_{0.1}O_3$.

**Table 1:** Results of Rietveld refinement of neutron diffraction pattern at 6K and 300K, Curie – Weiss fit parameters, geometrical frustration parameter, and transition temperature for  $YMn_{1-x}M_xO_3$ (M = Ti, Fe, Ga; x = 0, 0.1).

|  | $YMnO_3$ | | $YMn_{0.9}Ti_{0.1}O_3$ | | $YMn_{0.9}Fe_{0.1}O_3$ | | $YMn_{0.9}Ga_{0.1}O_3$ | |
|---|---|---|---|---|---|---|---|---|
|  | **6 K** | **300 K** | **6 K** | **300 K** | **6 K** | **300 K** | **6 K** | **300 K** |
| a (Å) | 6.1212(4) | 6.1402(2) | 6.1412(4) | 6.1579(3) | 6.1359(4) | 6.1510(4) | 6.1378(4) | 6.1549(4) |
| c (Å) | 11.4002(9) | 11.3901(8) | 11.3695(8) | 11.3605(8) | 11.4289(9) | 11.4117(17) | 11.3597(14) | 11.3560(12) |
| V (Å$^3$) | 369.93 | 371.90 | 371.35 | 373.07 | 372.64 | 373.91 | 370.61 | 372.56 |
| Mn-O1 (Å) | 1.90 (2) | 1.91 (2) | 1.94 (3) | 1.87 (7) | 1.88 (3) | 1.79 (5) | 1.90 (7) | 1.87 (5) |
| Mn-O2 (Å) | 1.86 (2) | 1.84 (2) | 1.79(3) | 1.87 (7) | 1.82 (3) | 1.94 (5) | 1.81 (7) | 1.87 (5) |
| Mn-O3 (Å) | 2.082 (3) | 2.09 (3) | 2.12 (3) | 2.12 (7) | 2.08 (2) | 1.98 (5) | 2.07 (6) | 2.11 (6) |
| Mn-O4 (Å) | 2.039 (3) | 2.042 (15) | 2.023 (15) | 2.03 (3) | 2.043 (16) | 2.10 (3) | 2.05 (3) | 2.03 (3) |
| Y1-O1(×3) (Å) | 2.301(20) | 2.28(2) | 2.33(4) | 2.26(6) | 2.29(3) | 2.41(4) | 2.32(5) | 2.28(2) |
| Y1-O2(×3) (Å) | 2.286(12) | 2.315(16) | 2.29(2) | 2.31(3) | 2.33(2) | 2.29(2) | 2.31(4) | 2.35(2) |
| Y1-O3 (Å) | 2.30(4) | 2.32(6) | 2.25(7) | 2.23(10) | 2.39(7) | 2.36(7) | 2.36(10) | 2.37(7) |
| Y2-O1(×3) (Å) | 2.271(12) | 2.263(13) | 2.28(2) | 2.28(3) | 2.252(18) | 2.25(3) | 2.24(3) | 2.25(2) |
| Y2-O2(×3) (Å) | 2.276(17) | 2.294(17) | 2.28(2) | 2.32(5) | 2.33(3) | 2.28(4) | 2.30(5) | 2.30(3) |
| Y2 - O4 (Å) | 2.45(3) | 2.46(3) | 2.58(6) | 2.56(10) | 2.46(6) | 2.48(6) | 2.44(8) | 2.49(6) |
| Mn-O3-Mn (°) | 119.24 (12) | 119.1 (6) | 119.4(6) | 118 (4) | 119 (2) | 119.3 (10) | 119 (4) | 119 (4) |
| Mn-O4-Mn (°) | 118.51 (11) | 118.8(7) | 119.2(6) | 120.0(14) | 119.1(6) | 119.3 (10) | 118.5(11) | 119.6 (11) |
| θ (K) | - 421 | | -119 | | -334 | | -382 | |
| $\mu_{eff}$ ($\mu_B$) | 4.98 | | 4.30 | | 4.45 | | 4.10 | |
| $\chi_0$ (emu/mol Oe) | 0.001 | | 0.001 | | 0.001 | | 0.0004 | |
| f=(θ/$T_N$) | 5.6 | | 2.2 | | 5. 6 | | 6.9 | |
| $T_N$ (K) | 75 | | 55 | | 60 | | 55 | |

**Table2:** Mössbauer parameters at room temperature for $YMn_{0.9}Fe_{0.1}O_3$.

| x | Iron Sites | [a]Isomer shift ($\delta$) mm/s | Quadrupole splitting ($\Delta E_Q$) mm/s | Line width ($\Gamma$) mm/s | Relative Area $R_A$ (%) |
|---|---|---|---|---|---|
| | Doublet 1 | 0.303 ± 0.002 | 1.940 ± 0.004 | 0.256±0.012 | 53.55 |
| 0.1 | Doublet 2 | 0.303 ± 0.004 | 2.109 ± 0.008 | 0.23 ±0.00 | 27.55 |
| | Doublet 3 | 0.289 ± 0.014 | 1.295 ± 0.028 | 0.446±0.038 | 18.90 |

[a]Isomer shift values are relative to Fe metal foil.

Fig 1

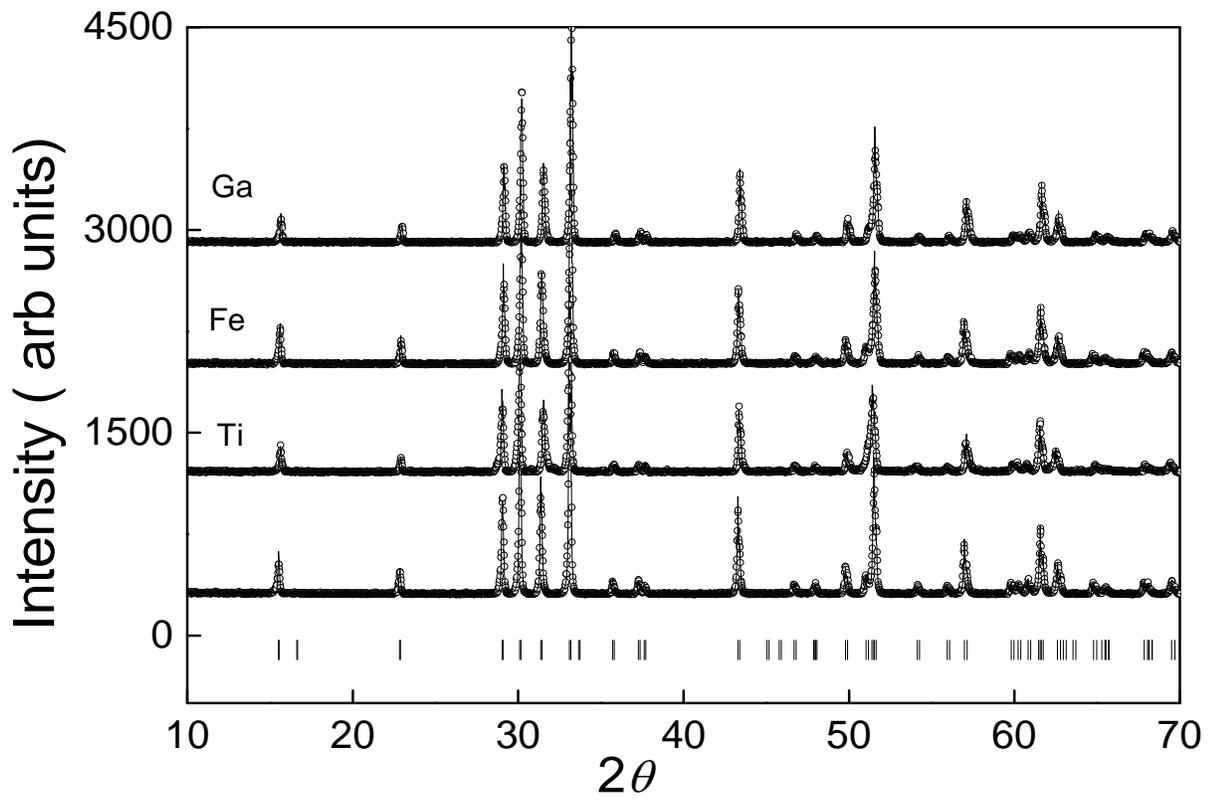

Fig 2

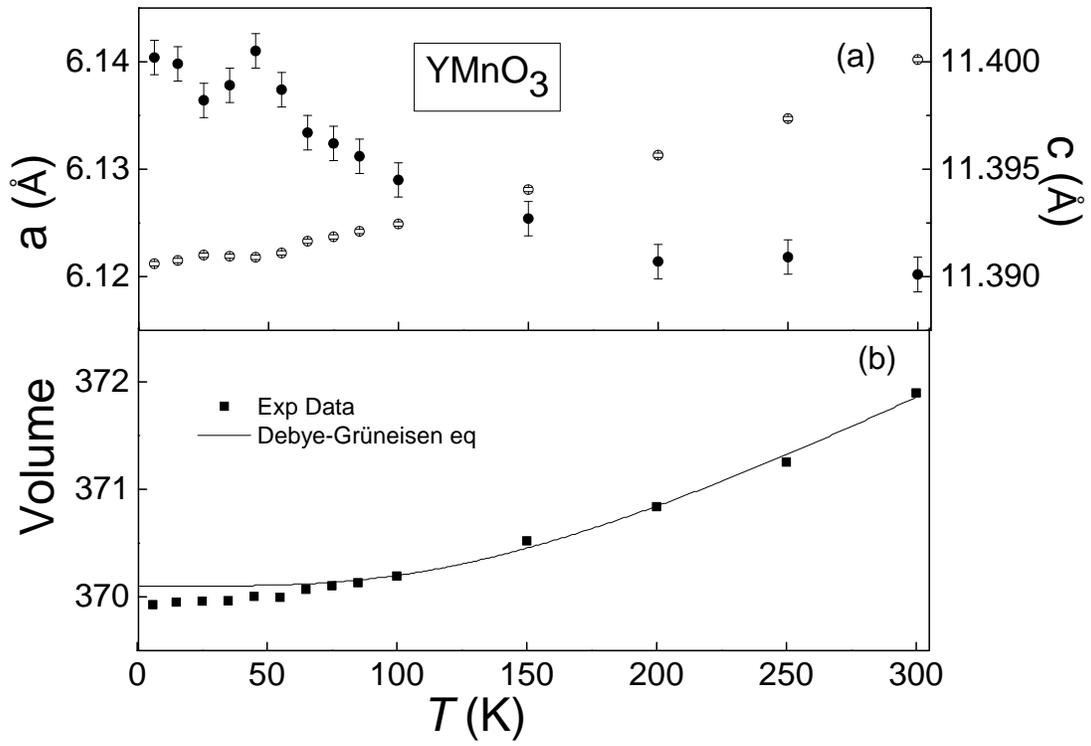

Fig 3

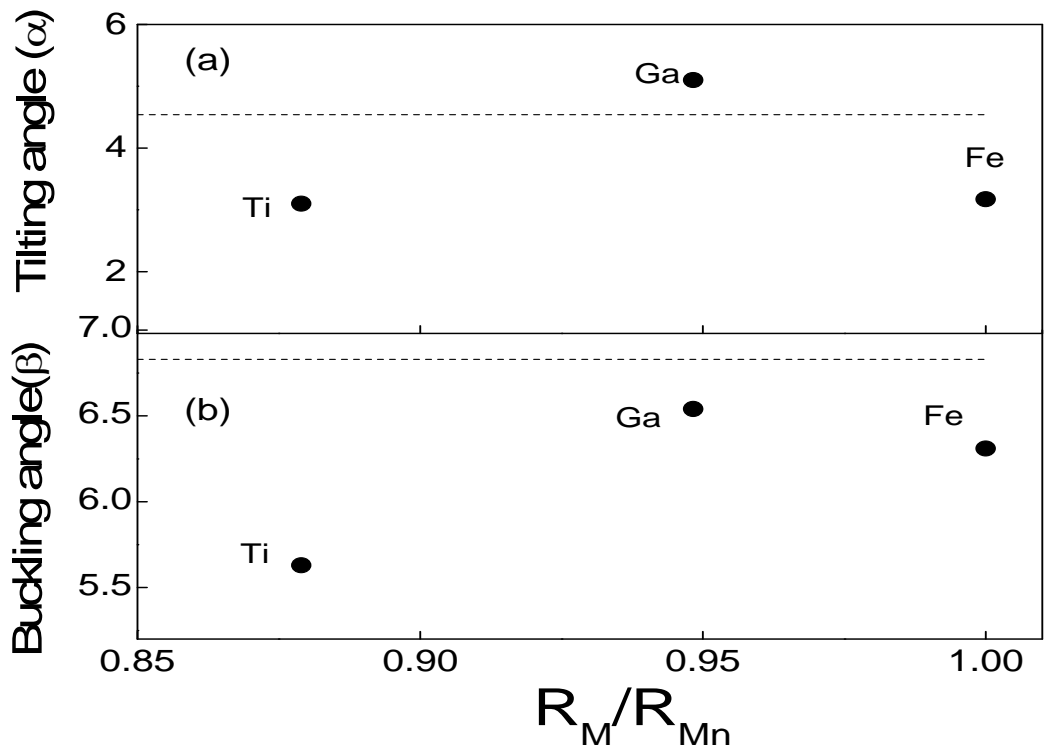

Fig 4

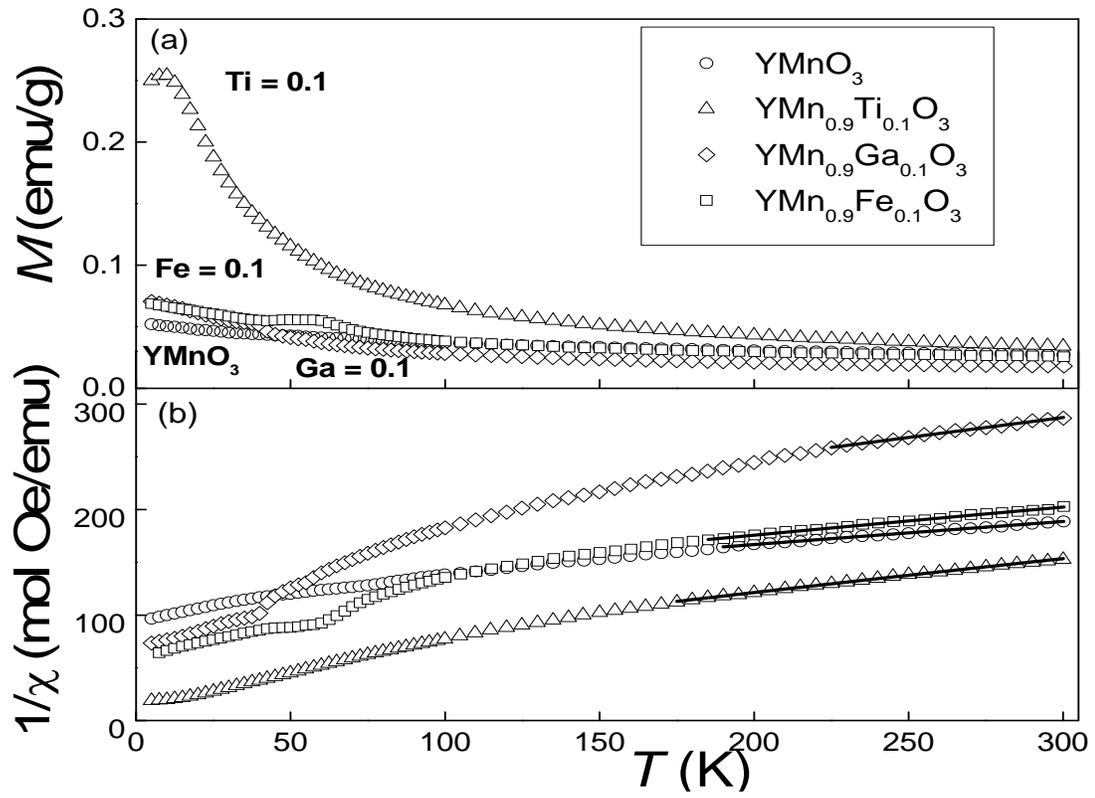

Fig 5

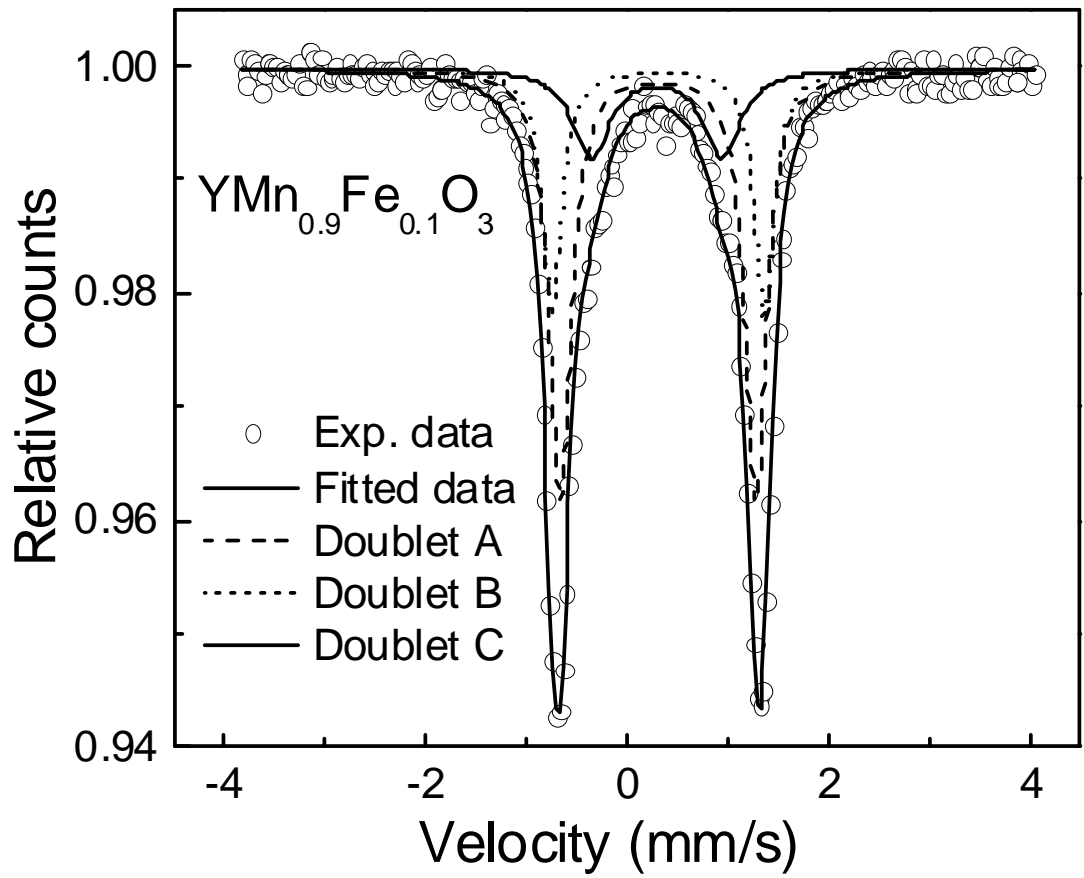

Fig 6

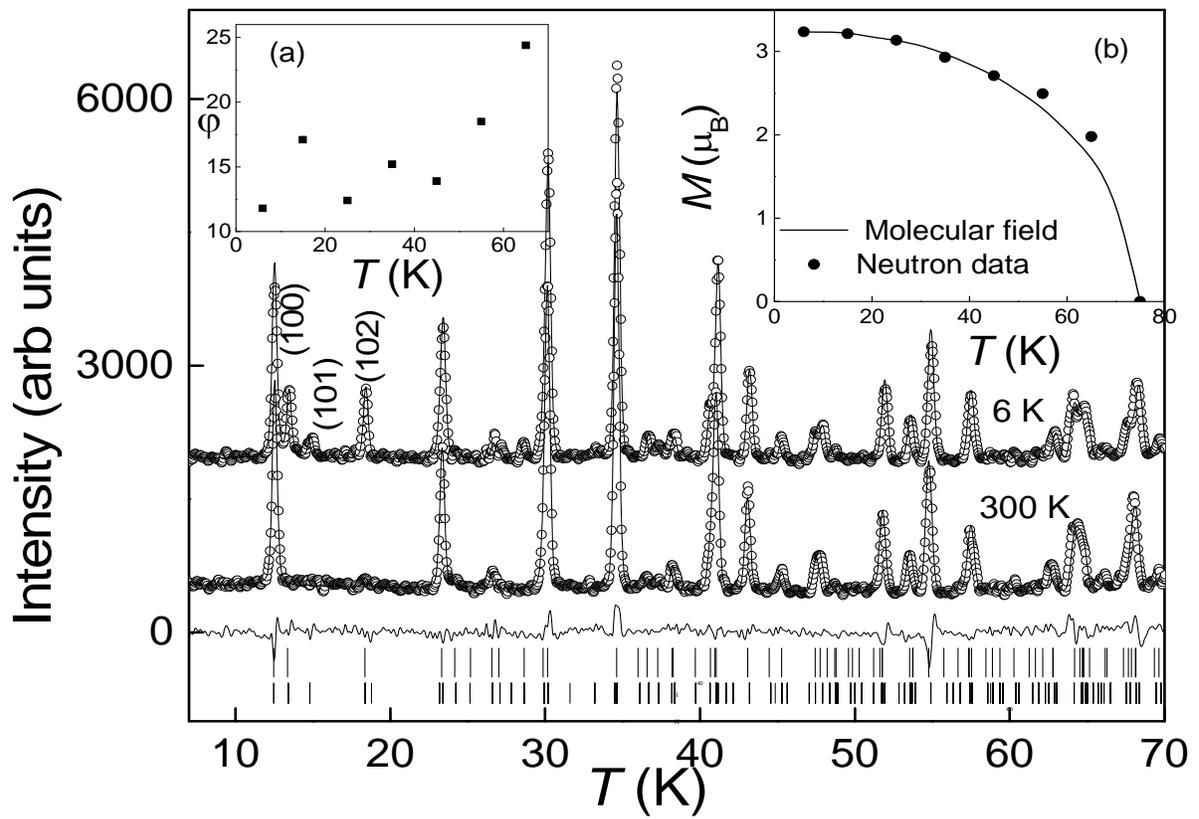

Fig 7

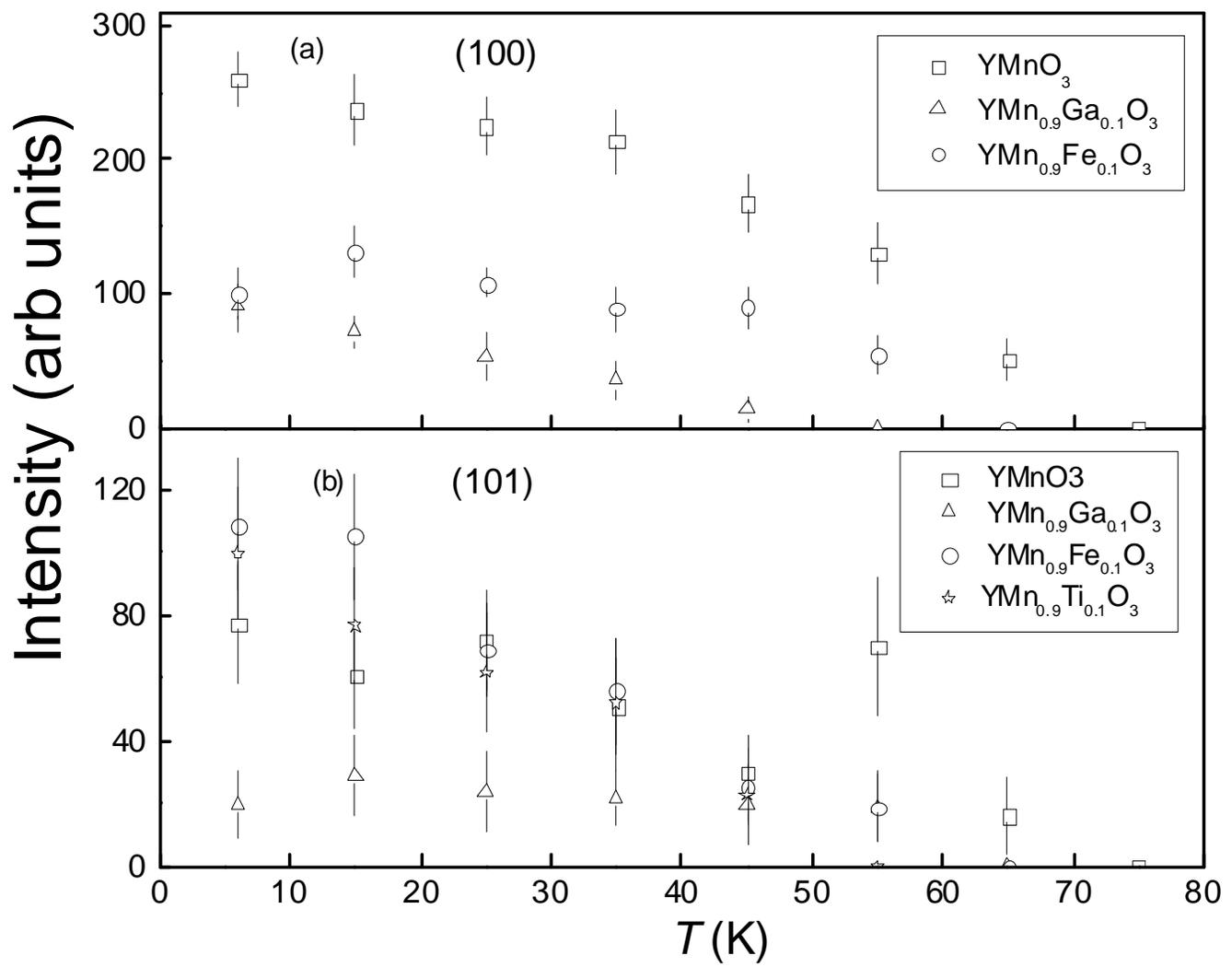

Fig 8

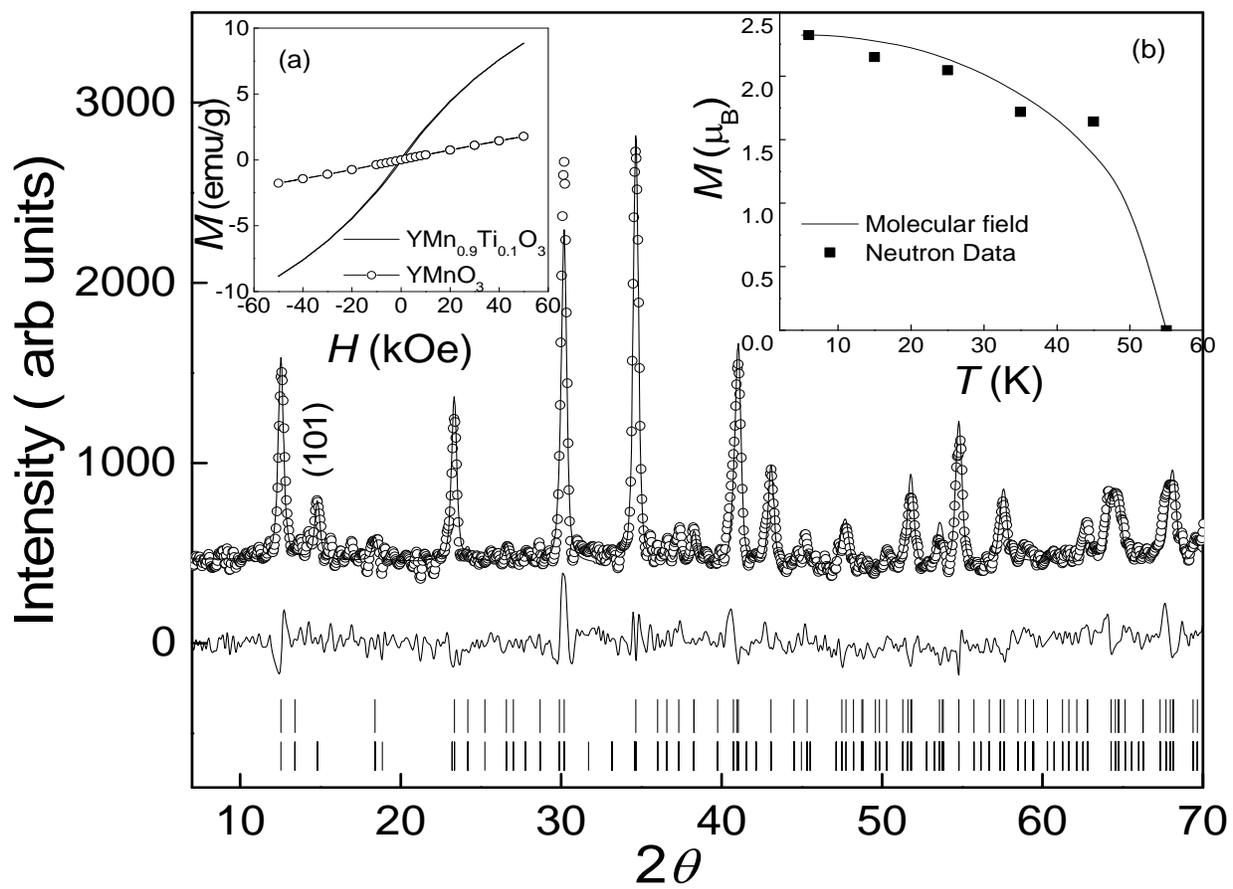

Fig 9

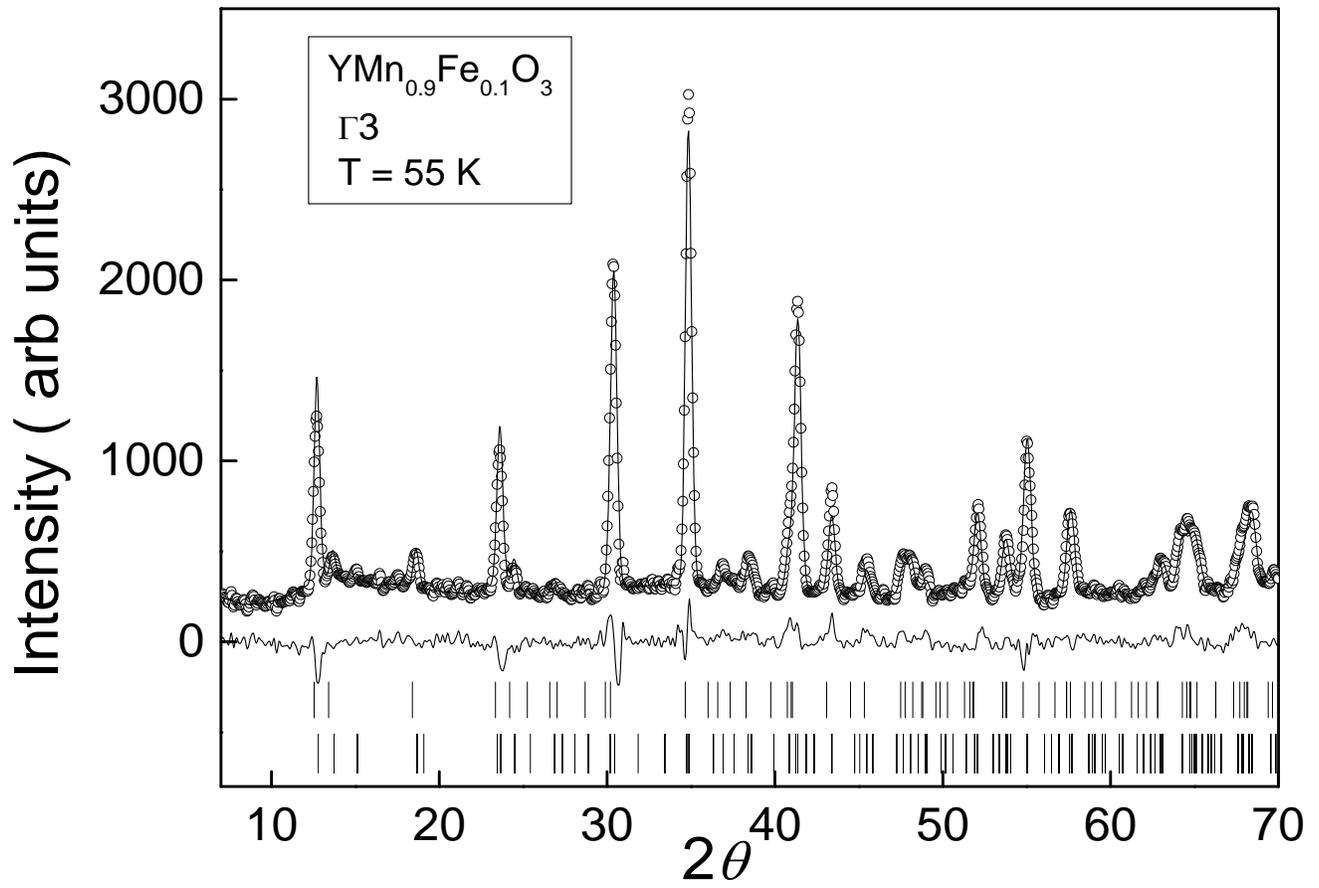

Fig 10

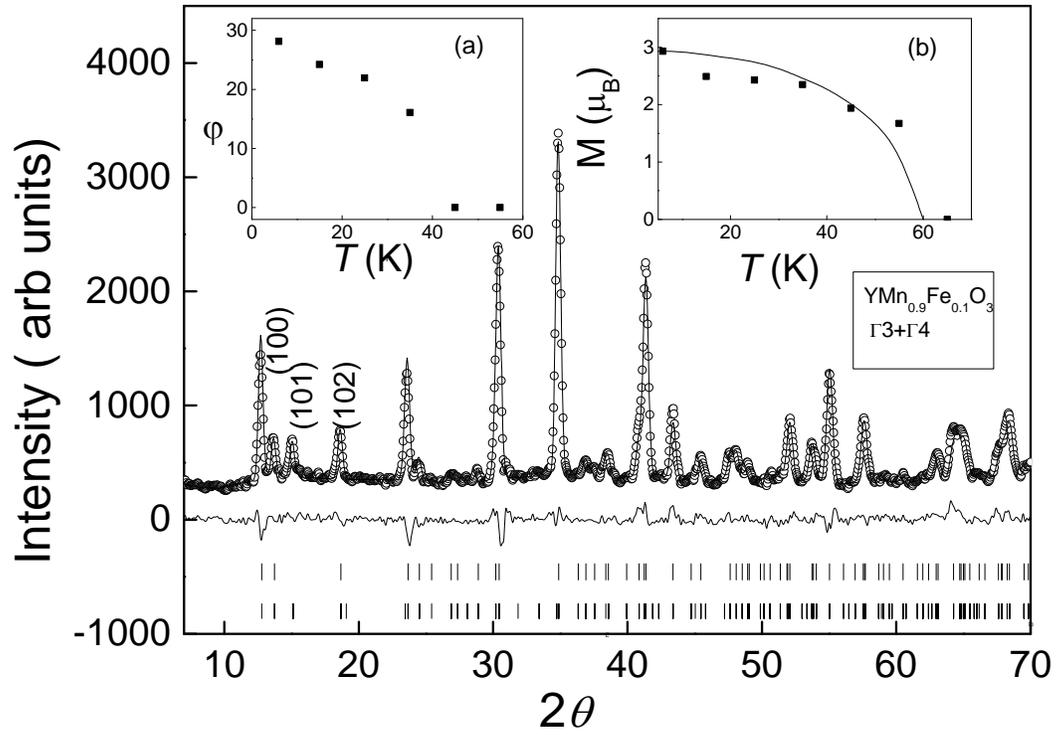